\documentclass[11pt]{article}
\usepackage[final]{acl}

\usepackage{times}
\usepackage{latexsym}

\usepackage[T1]{fontenc}

\usepackage[utf8]{inputenc}

\usepackage{microtype}

\usepackage{inconsolata}

\usepackage{graphicx}

\usepackage{algorithm}
\usepackage{amsmath}
\usepackage{booktabs}
\usepackage{multirow}
\usepackage{amssymb}
\usepackage{algpseudocode}
\usepackage{xcolor}
\usepackage{subcaption}
\usepackage{tabularx}
\usepackage{makecell}
\usepackage{adjustbox}
\usepackage{arydshln}

\title{RosettaSpeech: Zero-Shot Speech-to-Speech Translation \\without Parallel Speech}

\author{
\textbf{Zhisheng Zheng}\textsuperscript{1,2}\thanks{Work done during internship at Amazon.},
\textbf{Xiaohang Sun}\textsuperscript{2},
\textbf{Tuan Dinh}\textsuperscript{2},
\textbf{Abhishek Yanamandra}\textsuperscript{2},\\
\textbf{Abhinav Jain}\textsuperscript{2},
\textbf{Zhu Liu}\textsuperscript{2},
\textbf{Sunil Hadap}\textsuperscript{2},
\textbf{Vimal Bhat}\textsuperscript{2}, 
\textbf{Manoj Aggarwal}\textsuperscript{2},\\
\textbf{Gerard Medioni}\textsuperscript{2}, 
\textbf{David Harwath}\textsuperscript{1}\thanks{Corresponding author}
\\\\
\textsuperscript{1}University of Texas at Austin\qquad
\textsuperscript{2}Amazon\\ 
}

\begin{document}
\maketitle
\begin{abstract}

End-to-end speech-to-speech translation (S2ST) systems typically struggle with a critical data bottleneck: the scarcity of parallel speech-to-speech corpora. To overcome this, we introduce \textbf{RosettaSpeech}, a novel zero-shot framework trained exclusively on monolingual speech-text data augmented by machine translation supervision. Unlike prior works that rely on complex cascaded pseudo-labeling, our approach strategically utilizes text as a semantic bridge during training to synthesize translation targets, thereby eliminating the need for parallel speech pairs while maintaining a direct, end-to-end inference pipeline. Empirical evaluations on the CVSS-C benchmark demonstrate that RosettaSpeech achieves state-of-the-art zero-shot performance, surpassing leading baselines by significant margins—achieving ASR-BLEU scores of 25.17 for German-to-English (+27\% relative gain) and 29.86 for Spanish-to-English (+14\%). Crucially, our model effectively preserves the source speaker's voice without ever seeing paired speech data. We further analyze the impact of data scaling and demonstrate the model's capability in many-to-one translation, offering a scalable solution for extending high-quality S2ST to ``text-rich, speech-poor'' languages. 

\end{abstract}

\section{Introduction}
Speech-to-speech translation (S2ST) stands as a pivotal technology for dismantling language barriers, enabling seamless and natural communication across the globe. The ultimate goal is to create systems that can not only accurately translate spoken content but also preserve the rich paralinguistic information of the original speaker—such as tone, emotion, and prosody—in real-time.

Conventional approaches tackle this challenge with cascaded systems, which chain together separate Automatic Speech Recognition (ASR), Machine Translation (MT), and Text-to-Speech (TTS) models~\cite{nakamura2006atr, wahlster2013verbmobil}. While this modular design benefits from leveraging highly optimized, pre-trained components, it suffers from critical limitations, including error propagation, significant latency, and a fundamental inability to transfer prosodic information.

To address these issues, end-to-end (E2E) models have emerged, offering a direct mapping from source to target speech within a single neural network~\cite{jia2019direct, jia2022translatotron, zhang2024streamspeech}. These E2E systems can mitigate latency but struggle to effectively preserve the source speaker's voice, and their development is severely hampered by a data bottleneck: they require massive, parallel speech-to-speech translation (S2ST) corpora, which are prohibitively expensive and exist for only a handful of high-resource languages. Recent efforts in unsupervised S2ST have sought to overcome this data scarcity by using only monolingual data. However, these methods often rely on complex, multi-stage training pipelines, pseudo-labeling from cascaded models~\cite{wang2022simple}, or specialized architectures that can be difficult to train and scale~\cite{barrault2023seamlessm4t, fang2024can, nachmani2024translatotron}.


In this work, we introduce RosettaSpeech, a novel and simplified framework for zero-shot\footnote{We define ``zero-shot'' in this context as the complete absence of parallel \textit{source-speech} to \textit{target-speech} data.} speech-to-speech translation. Unlike prior works that rely on complex pipelines, our approach bridges the modality gap by leveraging off-the-shelf NMT models to transform abundant monolingual speech-text pairs into synthetic S2ST training targets. Although this implies a dependency on text-based translation resources, it successfully circumvents the critical bottleneck of acquiring expensive parallel speech corpora.

Crucially, RosettaSpeech is designed to address the "\textit{asymmetric resource}" scenario prevalent in many world languages. While thousands of languages have achieved a "text digitization milestone"—possessing decent text translation models or bitexts—they lack the massive parallel speech data required for conventional S2ST. By decoupling the need for speech parallelism from linguistic supervision, our framework offers a scalable path to unlock high-quality, speaker-preserving S2ST for this broad array of "\textit{text-rich, speech-poor}" languages.

Our primary contributions are as follows:
\begin{enumerate}
    \item We propose RosettaSpeech, a framework that fundamentally redefines S2ST training by utilizing text as a semantic bridge to exploit abundant monolingual data. By synthesizing translation targets via NMT, our approach achieves state-of-the-art results on standard benchmarks without requiring any \textit{parallel speech-to-speech corpora}.
    \item We demonstrate that a single model can be trained to perform many-to-one translation (e.g., French/Spanish/German to English) and achieve exceptional performance.
    \item We provide a foundational analysis of how training data and steps affect model performance, providing a crucial empirical foundation for future work on scaling S2ST systems to even larger models and more languages.
\end{enumerate}

By demonstrating that high-quality, speaker-preserving S2ST is achievable without parallel speech corpora, RosettaSpeech paves the way for developing powerful translation systems for a much wider and more diverse set of the world's languages.

\section{Related Work}
\subsection{Cascaded Speech Translation}
Conventional approaches to Speech-to-Speech Translation (S2ST) employ a cascaded architecture, decomposing the task into a sequence of three independently optimized modules~\cite{jain1991connectionist, nakamura2006atr, wahlster2013verbmobil}: 1) Automatic Speech Recognition (ASR)~\cite{radford2023robust, baevski2020wav2vec, hsu2021hubert} to convert source speech to text, 2) Machine Translation (MT)~\cite{kudugunta2023madlad, cheng2025seed} to translate the source text to the target language, and 3) Text-to-Speech (TTS) synthesis~\cite{chen2025f5, du2024cosyvoice, peng2024voicecraft, zheng2025voicecraft, jia2025ditar} to generate the final audio output.
$$
\text{Speech}_{\text{src}} \xrightarrow{\text{ASR}} \text{Text}_{\text{src}} \xrightarrow{\text{MT}} \text{Text}_{\text{tgt}} \xrightarrow{\text{TTS}} \text{Speech}_{\text{tgt}}
$$
The primary advantage of this modular design is its simplicity and the ability to leverage powerful, pre-trained models for each sub-task. Each component can be developed and improved independently. However, this pipeline architecture introduces several critical limitations. A principal issue is \textit{error propagation}, where inaccuracies from the ASR model are passed to and often amplified by the subsequent MT component, degrading the final translation quality. Furthermore, the sequential processing of the three stages inherently introduces significant latency, rendering the system unsuitable for real-time communication. Perhaps most critically, the reliance on an intermediate text representation severs the transfer of \textit{paralinguistic information}—such as prosody, emotion, and speaker identity—from the source speech. This results in synthesized output that lacks the naturalness and expressiveness of the original speaker. While some studies~\cite{bahar2019comparative, wang2020fairseq} show that highly optimized cascaded systems can achieve competitive translation accuracy, they fundamentally struggle to overcome the challenges of high latency and the loss of prosodic fidelity.

\subsection{End-to-End Speech Translation}
\subsubsection{Supervised}
To overcome the inherent limitations of cascaded systems, research has increasingly shifted towards end-to-end (E2E) models~\cite{jia2019direct, jia2022translatotron, barrault2023seamlessm4t}. These models are designed to perform direct speech-to-speech translation within a single, jointly optimized neural network. By unifying the process, E2E systems can theoretically eliminate the latency introduced by sequential processing and mitigate the problem of error propagation. More importantly, this direct mapping from source to target speech allows for the preservation of paralinguistic information, enabling the translated audio to retain the prosody, emotion, and vocal characteristics of the original speaker.


Despite their conceptual advantages, the widespread adoption of supervised E2E models is severely constrained by a critical bottleneck: the scarcity of parallel S2ST data. Training these large, data-hungry networks requires extensive corpora where the same utterance is available in both the source and target languages, often spoken by the same individual to maintain vocal consistency. Creating such datasets is exceptionally expensive and time-consuming, meaning they exist for only a handful of high-resource language pairs. This fundamental data dependency remains the primary obstacle to scaling E2E solutions to the vast majority of the world's languages.

\subsubsection{Unsupervised}
To overcome the scarcity of parallel data, unsupervised speech-to-speech translation (S2ST) leverages monolingual resources through innovative methods. Strategies range from creating pseudo-labels by cascading unsupervised ASR, MT, and TTS systems~\cite{wang2022simple}, to training direct end-to-end models like Translatotron 3~\cite{nachmani2024translatotron}, which uses back-translation and unsupervised embedding mapping. Further advancements focus on specific capabilities like expressivity and modularity. SONAR EXPRESSIVE~\cite{duquenne2023sonar} disentangles semantic content from a separately learned prosody embedding to enable zero-shot expressive translation, while ComSpeech~\cite{fang2024can} introduces a composite framework with a vocabulary adaptor and contrastive learning to combine arbitrary pretrained S2TT and TTS models, achieving high-quality S2ST without any parallel speech data. However, these methods often rely on complex, multi-stage pipelines or specialized model architectures, which can be difficult to train and scale effectively across many languages.

\subsection{Omni-Language Models} 
Omni-Language Models~\cite{xie2024mini, fang2024llama, chen2024slam, defossez2024moshi} are naturally suited for speech-to-speech related tasks. However, they are often trained on specialized and expensive data like speech-based question-answering (QA) corpora, which limits scalability. RosettaSpeech bypasses this bottleneck by training exclusively on abundant, monolingual speech-text pairs, requiring no parallel S2S or QA data. This data-efficient approach simplifies the training pipeline and makes it feasible to build S2ST systems for a much wider range of languages.

\section{Methodology}
\subsection{Datasets}

A core challenge in developing robust speech-to-speech translation (S2ST) systems is the profound scarcity of true parallel corpora, which contain the same utterance spoken in both a source and target language. Acquiring such data is prohibitively expensive and labor-intensive. 

Our methodology addresses this by explicitly decoupling speech supervision from translation supervision. We prioritize reliance on NMT over TTS-dependent pipelines for a strategic reason: parallel text data is orders of magnitude more abundant and accessible than the high-fidelity, studio-grade speech corpora required to train robust TTS systems. By utilizing NMT to generate pseudo-parallel targets, we effectively trade the difficult constraint of acquiring paired speech for the much looser constraint of acquiring parallel text.

The construction process involves two primary workflows, depending on the language of the monolingual source data:
\begin{enumerate}
    \item \textbf{From Source Language Data}: For a given monolingual corpus in the source language, which provides $(S_{src}, T_{src})$ pairs, we use a high-quality neural machine translation (NMT) model\footnote{\url{https://huggingface.co/google/madlad400-3b-mt}}~\cite{kudugunta2023madlad} to translate the source text $T_{src}$ into the target language, creating $T_{tgt}$. This procedure yields a pseudo-parallel triplet of $(S_{src}, T_{src}, T_{tgt}^{*})$.
    \item \textbf{From Target Language Data}: Conversely, for a monolingual corpus in the target language providing $(S_{tgt}, T_{tgt})$ pairs, we apply the same NMT model in the reverse direction to translate $T_{tgt}$ into $T_{src}$. This results in a corresponding triplet formatted as $(T_{src}^{*}, T_{tgt}, S_{tgt})$.
\end{enumerate}

Our strategy repurposes a diverse set of existing datasets to create a pseudo-parallel corpus. To ensure its quality and fidelity, we apply a rigorous filtering process. After generating translations, we use the COMET~\cite{rei2020comet} metric to score the quality of each source-target $T_{src}$-$T_{tgt}$ pair. Only pairs that meet or exceed a predefined threshold (e.g., 0.80 for EN$\leftrightarrow$DE) are retained for training. This quality control mechanism is critical, as it prevents the model from learning from inaccurate translations and thereby enhances the final system's performance and reliability. The resulting curated corpus provides the structured data needed to train our model to map from a source modality (speech or text) to target text and speech tokens, all without requiring direct $(S_{src}, S_{tgt})$ parallel pairs.

\begin{figure}[t]
    \centering
    \includegraphics[scale=0.48]{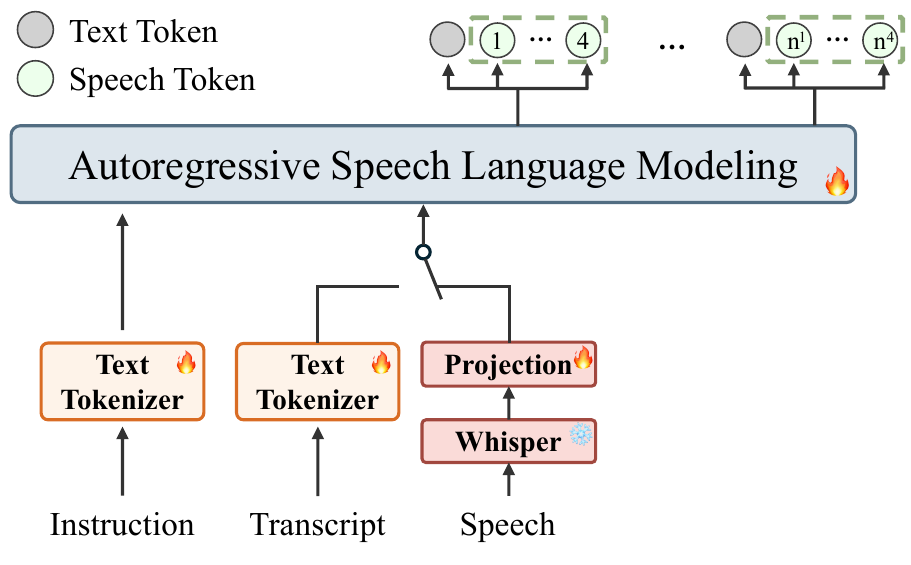}
    \caption{The training architecture of RosettaSpeech.}
    \label{fig:architecture}
    \vspace{-0.5cm}
\end{figure}
\subsection{Architecture}
As shown in Figure~\ref{fig:architecture}, the architecture of RosettaSpeech is designed to process source speech to generate translated text and speech. This section details the model's core components: speech modeling, the LLM backbone, and the multi-head projection layers.

\subsubsection{Speech Modeling}
Our approach processes input speech using the encoder from Whisper-medium~\cite{radford2023robust}. It transforms the raw 16kHz audio waveform into a sequence of continuous hidden-state vectors. These vectors are generated at a rate of 50 Hz and encapsulate rich acoustic and phonetic features of the source speech. For the output, the target speech waveform is converted into a sequence of discrete semantic tokens using the speech tokenizer from CosyVoice2~\cite{du2024cosyvoice}. This tokenized representation enables the LLM to generate speech autoregressively by predicting the next token in the sequence.

\subsubsection{Backbone}
The core of our architecture is the Qwen3-0.6B Large Language Model (LLM)~\cite{yang2025qwen3}, which functions as the central engine for all translation tasks. The selection of this model was primarily motivated by its advanced capabilities in text understanding. Critically, its inherent multilingual proficiency is essential for the cross-lingual objectives of our work. By leveraging a pre-trained LLM as the backbone, RosettaSpeech inherits the extensive world knowledge and complex linguistic patterns acquired during the LLM's foundational training, thereby providing a robust starting point for our specialized translation tasks.

\subsubsection{Multi-Head Projection}
To enable the model to generate both text and speech outputs, we employ a multi-head projection mechanism on top of the LLM backbone~\cite{xie2024mini, chen2024slam}. The final hidden-state representations from the LLM are passed to separate linear projection heads for each modality:

\begin{itemize}
\item A text projection head maps the hidden states to the vocabulary of the text tokenizer. This head is responsible for predicting the target text translation.
\item A set of speech projection heads maps the same hidden states to the vocabularies of the discrete speech tokenizer's codebooks. These heads work in concert to predict the sequence of semantic tokens required for synthesizing the target speech.
\end{itemize}

This multi-head design allows the model to be trained jointly on both text and speech generation tasks, learning to produce aligned outputs in both modalities from a shared latent representation.

\subsection{Training}
The model is trained following the procedure detailed in Algorithm~\ref{alg:pesudo-code-training}. The training process alternates between two distinct tasks based on the data sampled in each iteration:

\begin{enumerate}
    \item \textbf{Speech-to-Text Translation (S2TT):} This branch utilizes a monolingual source corpus $(S_{src}, T_{src})$. The model is fed the speech input $S_{src}$ and is trained to generate its text translation, while the original transcript $T_{src}$ is disregarded. To supervise the training, we compute the loss against a pseudo-target translation $T_{tgt}^{*}$ generated by an NMT system. Consequently, only the model's text output is used, and any generated speech is discarded.

    \item \textbf{Text-to-Speech Translation (T2ST)}: For data from the target language monolingual corpus $(S_{tgt}, T_{tgt})$, we use the NMT-generated pseudo-source text $T_{src}^{*}$ as input. The model is then trained to produce both the ground-truth target text $T_{tgt}$ and the ground-truth target speech $S_{tgt}$. The loss is calculated for both modalities.
\end{enumerate}

For both text and speech generation, the training objective is to minimize the standard cross-entropy loss between the model's predictions and the target sequences. The total loss for a given training batch is the sum of the S2TT and T2ST losses:
$$\mathcal{L}_{\text{total}} = \mathcal{L}_{\text{S2TT}} + \mathcal{L}_{\text{T2ST}}$$
where each loss term is defined as:
$$\mathcal{L} = -\sum_{i=1}^{N} \log P(y_i | y_{<i}, X; \theta)$$
Here, $X$ is the input sequence, $y_i$ is the $i$-th token of the target sequence (either text or discrete speech tokens), and $\theta$ represents the model parameters. This joint training approach enables the model to learn mappings from both speech and text modalities to a shared latent space that can produce aligned text and speech outputs.

\subsection{Inference}
Although the model has never been exposed to any paired source and target speech data $(S_{\text{src}}, S_{\text{tgt}})$ during training, it can perform direct speech-to-speech translation at inference time. The model takes the source speech waveform, $S_{\text{src}}$, as input and autoregressively generates both the translated text, $T_{\text{tgt}}$, and the translated semantic speech tokens, $S_{\text{tgt}}$, in a single forward pass. We then employ an off-the-shelf conditional flow matching (CFM) model~\cite{du2024cosyvoice} to convert these semantic tokens into mel-spectrograms. Crucially, this stage allows for effective control over the speaker identity and paralinguistic information in the synthesized audio by conditioning the generation on a given speech prompt. Finally, the mel-spectrograms are synthesized into the output waveform using a HiFi-GAN vocoder~\cite{kong2020hifi}.

\begin{algorithm}[t]
\caption{Pipeline of RosettaSpeech}
\label{alg:pesudo-code-training}
\begin{algorithmic}[1]
    \Require
    [I] Monolingual source corpus $\mathcal{D}_{\text{src}}$ and target corpus $\mathcal{D}_{\text{tgt}}$\;
    [II] Pre-trained neural machine translation models $\text{NMT}_{\text{src} \to \text{tgt}}$ and $\text{NMT}_{\text{tgt} \to \text{src}}$
    \Ensure A model $f_\theta$ for direct and zero-shot speech-to-speech translation $S_{\text{src}} \to S_{\text{tgt}}$
    \Statex 

    \Statex \textbf{Training}
    \For{each training iteration}
        \Statex \hspace{\algorithmicindent} \textit{// 1. Speech-to-Text Translation (S2TT)}
        \State Sample $(S_{\text{src}}, T_{\text{src}})$ from $\mathcal{D}_{\text{src}}$
        \State $T_{\text{tgt}}^{*} \leftarrow \text{NMT}_{\text{src} \to \text{tgt}}(T_{\text{src}})$
        \State $\hat{T}_{\text{tgt}}, \_ \leftarrow f_{\theta}(S_{\text{src}})$
        \State $\mathcal{L}_{\text{S2TT}} \leftarrow \text{Loss}(\hat{T}_{\text{tgt}}, T_{\text{tgt}}^{*})$
        \Statex 

        \Statex \hspace{\algorithmicindent} \textit{// 2. Text-to-Speech Translation (T2ST)}
        \State Sample $(S_{\text{tgt}}, T_{\text{tgt}})$ from $\mathcal{D}_{\text{tgt}}$
        \State $T_{\text{src}}^{*} \leftarrow \text{NMT}_{\text{tgt} \to \text{src}}(T_{\text{tgt}})$
        \State $\hat{T}_{\text{tgt}}, \hat{S}_{\text{tgt}} \leftarrow f_{\theta}(T_{\text{src}}^{*})$
        \State $\mathcal{L}_{\text{T2ST}} \leftarrow \text{Loss}(\hat{T}_{\text{tgt}}, T_{\text{tgt}}) + \text{Loss}(\hat{S}_{\text{tgt}}, S_{\text{tgt}})$
        \Statex

        \Statex \hspace{\algorithmicindent} \textit{// 3. Parameter Update}
        \State $\mathcal{L}_{\text{total}} \leftarrow \mathcal{L}_{\text{S2TT}} + \mathcal{L}_{\text{T2ST}}$
        \State Update $\theta$ using $\nabla_\theta \mathcal{L}_{\text{total}}$
    \EndFor
    \Statex 

    \Statex \textbf{Inference}
    \Statex \Return $(T_{\text{tgt}}, S_{\text{tgt}}) = f_\theta(S_{\text{src}})$
\end{algorithmic}
\end{algorithm}

\begin{table*}[htb]
\vspace{-0.8cm}
\centering
\resizebox{\textwidth}{!}{
\begin{tabular}{@{}lcccccc@{}}
    \toprule
    \multirow{2}{*}{\textbf{Models}} & \multicolumn{2}{c}{\textbf{FR → EN}} & \multicolumn{2}{c}{\textbf{ES → EN}} & \multicolumn{2}{c}{\textbf{DE → EN}} \\
    \cmidrule(lr){2-3} \cmidrule(lr){4-5} \cmidrule(lr){6-7}
    & \textbf{BLEU (↑)} & \textbf{ASR-BLEU (↑)} & \textbf{BLEU (↑)} & \textbf{ASR-BLEU (↑)} & \textbf{BLEU (↑)} & \textbf{ASR-BLEU (↑)} \\
    \midrule
    \textcolor{gray}{Ground Truth} & \textcolor{gray}{-} & \textcolor{gray}{84.52} & \textcolor{gray}{-} & \textcolor{gray}{88.54} & \textcolor{gray}{-} & \textcolor{gray}{75.53} \\
    \textcolor{gray}{Qwen3-0.6B~\cite{yang2025qwen3}} & \textcolor{gray}{32.68} & \textcolor{gray}{-} & \textcolor{gray}{34.35} & \textcolor{gray}{-} & \textcolor{gray}{27.42} & \textcolor{gray}{-} \\
    \midrule
    \multicolumn{7}{c}{\textbf{Zero-Shot}} \\
    \addlinespace[0.2em]
    ComSpeech~\cite{fang2024can} & 30.72 & \textbf{28.15} & 26.51 & 24.80 & 19.41 & 18.16 \\
    Ours (Unparalleled) & \textbf{31.78} & 27.86 & \textbf{32.64} & \textbf{29.86} & \textbf{31.95} & \textbf{25.17} \\
    \midrule
    \multicolumn{7}{c}{\textbf{Non Zero-Shot}} \\
    \addlinespace[0.2em]
    Translatotron~\cite{jia2019direct} & - & 16.96 & - & 8.72 & - & 1.97 \\
    Translatotron 2~\cite{jia2022translatotron} & 28.82 & 25.49 (26.07) & 25.82 & 22.35 (22.93) & 18.66 & 16.24 (16.91) \\
    S2UT~\cite{lee2021direct} & - & 20.91 (22.23) & - & 16.94 (18.53) & - & 2.46 (2.99) \\
    UnitY~\cite{inaguma2022unity} & - & 26.90 (27.77) & - & 23.93 (24.95) & - & 18.19 (18.74) \\
    DASpeech~\cite{fang2023daspeech} & - & 25.03 & - & 21.37 & - & 16.14 \\
    StreamSpeech~\cite{zhang2024streamspeech} & 31.59 (32.60) & 27.58 (28.45) & 28.97 (30.35) & 26.16 (27.25) & 21.96 (23.36) & 19.72 (20.93) \\
    Hibiki~\cite{labiausse2025high} & - & 30.5 & - & - & - & - \\
    \hline
    Ours (Paralleled, Scratch) & 24.16 & 15.32 & 23.62 & 9.38 & 17.71 & 9.02 \\
    Ours (Paralleled, Fine-tuned) & 32.88 & 31.56 & \textbf{35.23} & \textbf{33.05} & \textbf{32.62} & \textbf{29.90} \\
    Ours (Paralleled, Fine-tuned)$^\dagger$ & \textbf{33.11} & \textbf{32.16} & 30.92 & 29.35 & 23.22 & 21.54 \\
    \bottomrule

\end{tabular}
}

\parbox{\linewidth}{\vspace{0.5ex}\footnotesize
\raggedright $^\dagger$Indicates a single model for all three language pairs. Other results correspond to models trained for each pair individually.}
\caption{Speech-to-speech translation performance on the CVSS-C test set for FR/ES/DE$\to$EN. We reorganize results into two blocks: \textbf{Zero-shot} (ComSpeech, RosettaSpeech (Unparalleled)) and \textbf{Non zero-shot} (all others). Setup follows StreamSpeech. Results without parentheses are from greedy search, while those in parentheses are from beam search (beam size 10).}
\label{tab:main-results}
\vspace{-0.5cm}
\end{table*}

\section{Experiments}
\subsection{Datasets}
We validate our method on speech-to-speech translation for three language pairs: French to English (FR$\to$EN), German to English (DE$\to$EN), and Spanish to English (ES$\to$EN). The training data is constructed entirely from monolingual corpora, without relying on any parallel speech-to-speech data. For the target language, \textit{English}, we utilize speech-text pairs from the Gigaspeech~\cite{chen2021gigaspeech} and the English portion of the Multilingual LibriSpeech (MLS)~\cite{pratap2020mls} datasets. For the source languages—\textit{French, German, and Spanish}—we use the corresponding language-specific subsets from the VoxPopuli~\cite{wang2021voxpopuli} and Multilingual LibriSpeech corpora. These monolingual datasets form the basis from which we generate the pseudo-parallel data required for training, as detailed in the preceding section.

\subsection{Implementation Details}
\subsubsection{Training} 
For our model, we use CosyVoice2's speech tokenizer, which has a single-layer, 6561-entry codebook and operates at 25 Hz for 16kHz speech. The outputs from the final Transformer layer of the Qwen model are then projected into five distinct linear layers: one for text tokens and the remaining four for speech tokens. The same configuration was used for training the model for each language pair. We employed the AdamW optimizer with a learning rate of $2 \times 10^{-3}$, $\beta_1 = 0.9$, $\beta_2 = 0.999$, an epsilon of $1 \times 10^{-6}$, and a weight decay of 0.01. A learning rate scheduler is utilized, featuring a linear warm-up for the initial 10K steps, followed by a linear decay for the remainder of the 100K total training steps. Gradient accumulation is performed over micro-batches. The training of each model was done on 8 NVIDIA A100 40GB GPUs.
\subsubsection{Inference} 
During inference, we use distinct decoding strategies for the generation of text and speech tokens. For the text translation, we utilize greedy search, which we found empirically to yield higher-quality results than nucleus sampling for this task. Conversely, for the generation of discrete speech tokens, we use nucleus sampling with a TopK of 20, a TopP of 0.8, and a temperature of 0.95. Furthermore, we apply a length penalty during speech generation to prevent the model from producing repetitive or non-terminating audio outputs.

\subsection{Evaluation}
For a fair comparison, we follow the evaluation protocol of StreamSpeech~\cite{zhang2024streamspeech} and evaluate our model on the CVSS-C~\cite{jia2022cvss} test set for French-, German-, and Spanish-to-English speech-to-speech translation (FR/DE/ES $\to$ EN). We report results along two axes: \textit{quality} and \textit{fidelity}. Quality is measured using BLEU, ASR-BLEU, COMET, and BLASER 2.0, while fidelity is evaluated with speaker similarity (SIM) and the naturalness mean opinion score (NMOS). Detailed descriptions of these metrics are provided in the Appendix §\ref{app:metric}. We compare RosettaSpeech against a comprehensive set of state-of-the-art S2ST models, including Translatotron~\cite{jia2019direct}, Translatotron 2~\cite{jia2022translatotron}, S2UT~\cite{lee2021direct}, UnitY~\cite{inaguma2022unity}, DASpeech~\cite{fang2023daspeech}, ComSpeech~\cite{fang2024can}, StreamSpeech~\cite{zhang2024streamspeech}, and Hibiki~\cite{labiausse2025high}.

\subsection{Results}
\paragraph{Translation Quality} As presented in Table~\ref{tab:main-results} and Table~\ref{tab:quality2}, our RosettaSpeech model, trained exclusively without parallel speech, establishes a new state-of-the-art for zero-shot speech-to-speech translation on the CVSS-C benchmark. It significantly outperforms prior systems, even those trained with parallel speech data.

\begin{table}[htb]
\centering
\setlength{\tabcolsep}{2.5pt} 
\resizebox{\columnwidth}{!}{
\begin{tabular}{@{}lcccccc@{}}
    \toprule
    \multirow{2}{*}{\textbf{Models}} & \multicolumn{2}{c}{\textbf{FR $\to$ EN}} & \multicolumn{2}{c}{\textbf{ES $\to$ EN}} & \multicolumn{2}{c}{\textbf{DE $\to$ EN}} \\
    \cmidrule(lr){2-3} \cmidrule(lr){4-5} \cmidrule(lr){6-7}
    & \textbf{COMET} & \textbf{BLASER 2.0} & \textbf{COMET} & \textbf{BLASER 2.0} & \textbf{COMET} & \textbf{BLASER 2.0} \\
    \midrule
    ComSpeech & 70.13 & 3.69 & 66.96 & 3.66 & 57.96 & 3.33 \\
    StreamSpeech & 76.66 & 3.95 & 74.80 & 3.93 & 65.51 & 3.61 \\
    \hline
    Ours (Unparalleled) & 77.46	& 3.95 & 80.24 & 4.11 & 78.63 & 3.93\\
    Ours (Paralleled, FT) & \textbf{78.97} & \textbf{4.04} & \textbf{82.05} & \textbf{4.17} & \textbf{79.65} & \textbf{4.03} \\
    \bottomrule
\end{tabular}
}
\caption{Evaluation on the CVSS-C test set using COMET and BLASER 2.0 metrics.}
\label{tab:quality2}
\vspace{-0.3cm}
\end{table}

For French-to-English translation, our model achieves a highly competitive ASR-BLEU score of 27.86 and a BLEU score of 31.78, surpassing strong baselines like StreamSpeech. In the Spanish-to-English task, RosettaSpeech establishes a new benchmark with an ASR-BLEU of 29.86, marking a relative improvement of over 14\% against the previous leading system. The performance gains are most pronounced for German-to-English, where our model achieves an ASR-BLEU of 25.17, representing a substantial relative improvement of over 27\%. Beyond standard lexical metrics, Table~\ref{tab:quality2} highlights our model's superior performance on advanced neural metrics (COMET and BLASER 2.0). On these metrics, our unparalleled model consistently outperforms the strong StreamSpeech baseline. Notably, for Spanish-to-English, we achieve a COMET score of 80.24 compared to the baseline's 74.80, and a BLASER 2.0 score of 4.11 compared to 3.93. The margin is even wider for German-to-English, where our model scores 78.63 on COMET against the baseline's 65.51. These results confirm that RosettaSpeech generates translations that are not only textually accurate but also achieve higher overall quality scores in latent representation spaces.

\paragraph{Translation Fidelity}
\begin{table}[htb]
\vspace{-0.55cm}
\centering
\setlength{\tabcolsep}{2.5pt} 
\resizebox{\columnwidth}{!}{
\begin{tabular}{@{}lcccccc@{}}
    \toprule
    \multirow{2}{*}{} & \multicolumn{2}{c}{\textbf{FR $\to$ EN}} & \multicolumn{2}{c}{\textbf{ES $\to$ EN}} & \multicolumn{2}{c}{\textbf{DE $\to$ EN}} \\
    \cmidrule(lr){2-3} \cmidrule(lr){4-5} \cmidrule(lr){6-7}
    & \textbf{SIM} & \textbf{NMOS} & \textbf{SIM} & \textbf{NMOS} & \textbf{SIM} & \textbf{NMOS} \\
    \midrule
    CVSS-C & 0.03 & 3.82 & 0.05 & 3.83 & 0.01 & \textbf{3.79} \\
    CVSS-T & 0.21 & 3.27 & 0.19 & 3.09 & 0.21 & 2.91 \\ \hline
    ComSpeech & 0.02 & 3.14 & 0.04 & 2.60 & 0.01 & 2.55 \\
    StreamSpeech & 0.03 & 3.53 & 0.05 & 3.36 & 0.03 & 3.42 \\
    Ours (Unparalleled) & \textbf{0.36} & 3.61 & \textbf{0.36} & 3.79 & \textbf{0.38} & 3.17 \\
    Ours (Paralleled, FT) & 0.34 & \textbf{3.88} & 0.34 & \textbf{3.87} & 0.36 & 3.70 \\
    \bottomrule
\end{tabular}
}
\caption{Speaker similarity and speech naturalness evaluation (SIM and NMOS) on the CVSS-C test set.}
\label{tab:sim}
\vspace{-0.35cm}
\end{table}

As shown in Table~\ref{tab:sim}, existing S2ST baselines such as ComSpeech and StreamSpeech yield negligible SIM scores ($<0.05$), indicating a substantial failure to retain the source speaker's identity. In contrast, RosettaSpeech demonstrates robust zero-shot voice cloning capabilities. Remarkably, our model produces SIM scores ($\approx0.36$) that significantly exceed even the reference targets provided by the CVSS dataset, suggesting that our generated speech preserves the source speaker's characteristics more effectively than the benchmark's own synthetic data. Furthermore, our model achieves the best NMOS scores on nearly all language pairs, attesting to the high fidelity, naturalness, and excellent acoustic quality of our generations.

\paragraph{Translation Speed}
As shown in Table~\ref{tab:efficiency}, the cascaded baseline suffers from high latency (RTF 1.04) due to its massive size (4.3B parameters). In contrast, RosettaSpeech reduces the parameter count by nearly \textbf{80\%} (to \textit{0.9B}) and improves inference speed by \textbf{2$\times$} (RTF \textbf{0.53}).
\begin{table}[t]
\centering
\resizebox{\columnwidth}{!}{%
\begin{tabular}{@{}llcc@{}}
\toprule
\textbf{System} & \textbf{Component} & \textbf{Params} & \textbf{RTF ($\downarrow$)} \\
\midrule
\multirow{4}{*}{Cascaded Baseline} & ASR (Whisper-medium) & 0.8B & 0.26 \\
 & MT (MADLAD-400-3B) & 3.0B & 0.14 \\
 & TTS (CosyVoice 2) & 0.5B & 0.64 \\
 \cmidrule(l){2-4}
 & \textit{Total} & 4.3B & 1.04 \\
\midrule
\textbf{RosettaSpeech (Ours)} & \textbf{End-to-End} & \textbf{0.9B} & \textbf{0.53} \\
\bottomrule
\end{tabular}
}
\caption{Model size and latency comparison. We measure the Real-Time Factor (RTF) on a single NVIDIA A100 GPU. }
\label{tab:efficiency}
\vspace{-0.5cm}
\end{table}

\subsection{Ablation Studies}
\paragraph{Multi-stage training}
Our RosettaSpeech model is trained from the outset using a joint training strategy, where data for speech-to-text translation (S2TT) and text-to-speech translation (T2ST) are mixed and learned simultaneously. To demonstrate the necessity of this approach, we conducted an ablation study comparing it against two sequential training methods, with results shown in Table~\ref{tab:multi-stage}. The sequential methods suffer from severe catastrophic forgetting. For instance, when the model is trained first on S2TT and then on T2ST, it forgets how to process speech inputs, causing its S2ST performance to plummet to a near-zero ASR-BLEU of 0.18. Conversely, training on T2ST first and then S2TT erases the model's speech generation capabilities, resulting in an ASR-BLEU of just 0.49. In stark contrast, our joint training method avoids this issue, achieving strong and balanced performance across all four tasks.

\begin{table}[htb]
\centering
\resizebox{\linewidth}{!}{%
\begin{tabular}{@{} lcccc @{}}
\toprule
 & \textbf{T $\to$ T} & \textbf{T $\to$ S} & \textbf{S $\to$ T} & \textbf{S $\to$ S} \\
\midrule
\textcolor{gray}{Ground Truth} & \textcolor{gray}{-} & \textcolor{gray}{84.52} & \textcolor{gray}{-} & \textcolor{gray}{-} \\
\textcolor{gray}{Qwen3-0.6B} & \textcolor{gray}{32.68} & \textcolor{gray}{-} & \textcolor{gray}{-} & \textcolor{gray}{-} \\
\midrule

\multicolumn{5}{@{}l}{\textit{Sequential Training}} \\
S2TT $\rightarrow$ T2ST  & 33.84 & 28.53 & 0.15 & 0.18 \\
T2ST $\rightarrow$ S2TT  & 31.44 & 1.95 & 30.97 & 0.49 \\
\midrule

\multicolumn{5}{@{}l}{\textit{Joint Training}} \\
S2TT + T2ST & \textbf{34.14} & \textbf{29.27} & \textbf{31.78} & \textbf{27.86} \\
\bottomrule
\end{tabular}
}
\caption{Translation performance on the CVSS-C (French to English) test set. \textbf{T} = Text, \textbf{S} = Speech.}
\label{tab:multi-stage}
\vspace{-0.3cm}
\end{table}

\paragraph{Fine-tuning}
To further probe the capabilities of our framework, we explored the impact of fine-tuning the pre-trained RosettaSpeech model on a limited amount of parallel speech-to-speech data (CVSS-T\footnote{The English speech was regenerated using CosyVoice 2.}). As detailed in Table~\ref{tab:main-results}, this fine-tuning stage yields a substantial performance boost across all language pairs. For example, the ASR-BLEU score for French-to-English translation improves from 27.86 to 31.56, while the German-to-English score rises from 25.17 to 29.90. This demonstrates that our monolingual pre-training strategy creates a powerful foundation that can be rapidly and effectively specialized with even a small quantity of supervised data.

Furthermore, we investigated the model's capacity for multilingual, many-to-one translation by fine-tuning a single model to handle French, Spanish, and German to English translation simultaneously. As shown in the final row of Table~\ref{tab:main-results}, this unified model maintains competitive performance, particularly improving on French-to-English (+0.6 ASR-BLEU). However, we observe a performance degradation for Spanish and German compared to their specialized counterparts. We attribute this to \textit{capacity dilution} and \textit{language interference}. Given the relatively compact size of the backbone (0.6B parameters), the model faces a bottleneck when simultaneously encoding the distinct acoustic and syntactic features of multiple languages. This effect is most pronounced for German, likely due to its greater linguistic distance from the Romance languages (French and Spanish), leading to negative transfer during joint optimization.

\paragraph{Training Steps and Data Scaling}
\begin{figure}
    \vspace{-0.5cm}
    \centering
        \includegraphics[width=.9\linewidth, clip, trim={2.3cm 0.5cm 2.3cm 0.5cm}]{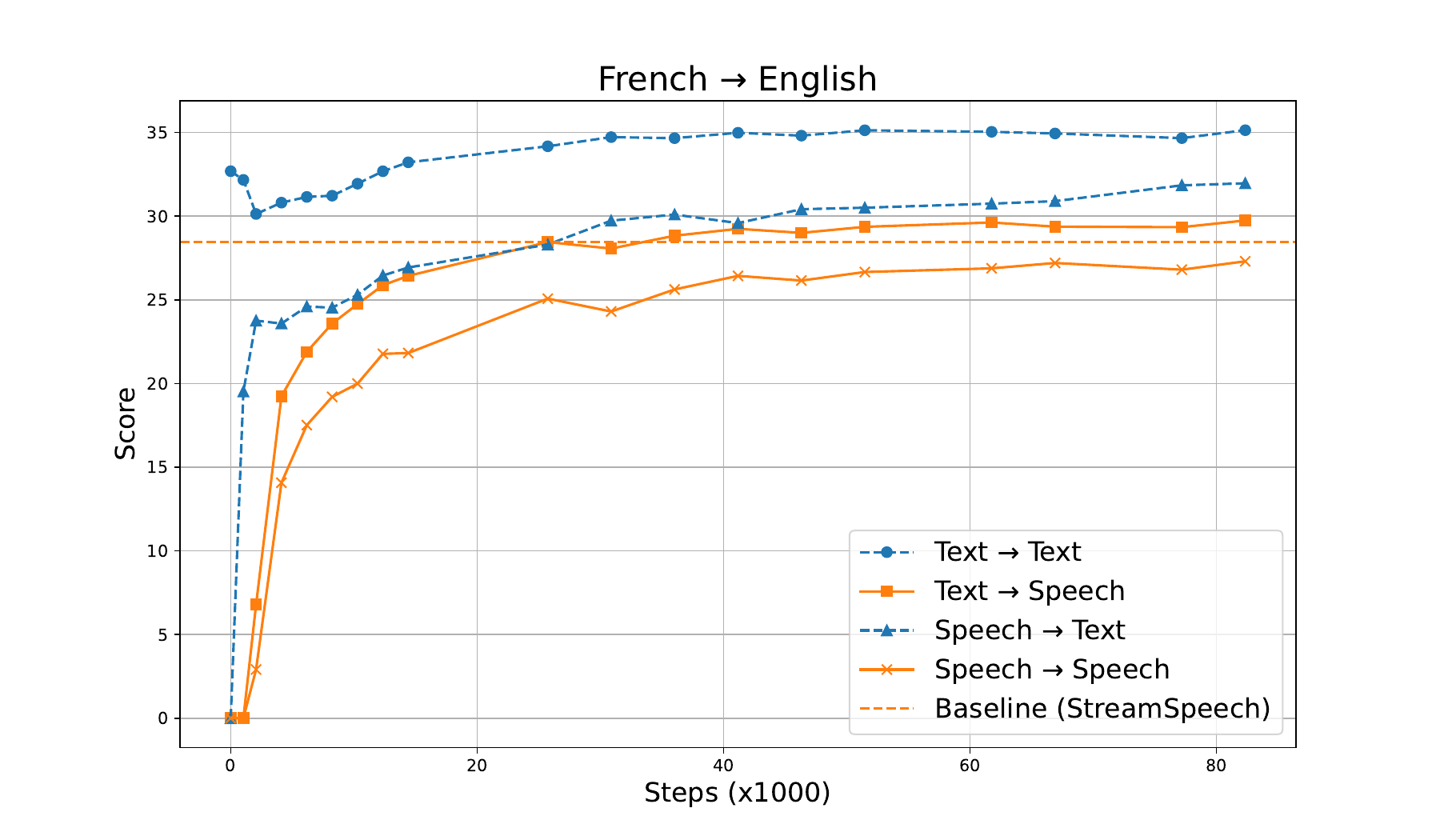}
    \caption{S2TT and S2ST results on the CVSS-C test set at different training steps, compared to the StreamSpeech baseline (the two dashed lines).}
    \label{fig:step_scale}
    \vspace{-0.3cm}
\end{figure}

To provide a foundational understanding of RosettaSpeech's behavior, we analyze its performance as a function of training steps and data volume.

Figure~\ref{fig:step_scale} illustrates the model's performance on the CVSS-C test set for French-to-English translation. It is worth noting that at the very beginning of training, we observe a slight, initial dip in text-to-text translation performance. This phenomenon is expected, as the pre-trained text-only backbone is being adapted to handle more complex, multi-modal objectives. However, as training progresses, this capability not only recovers but is further enhanced, leading to a clear and consistent trend: translation quality for all tasks—both text-output (Text$\to$Text, Speech$\to$Text) and speech-output (Text$\to$Speech, Speech$\to$Speech) — steadily improves with more training steps. Notably, our model surpasses the performance of the strong StreamSpeech baseline (indicated by the dashed orange line) relatively early in the training process, typically within the first 20,000 steps, highlighting its remarkable training efficiency. 

\begin{figure}[H]
    \centering
    \includegraphics[width=0.75\linewidth]{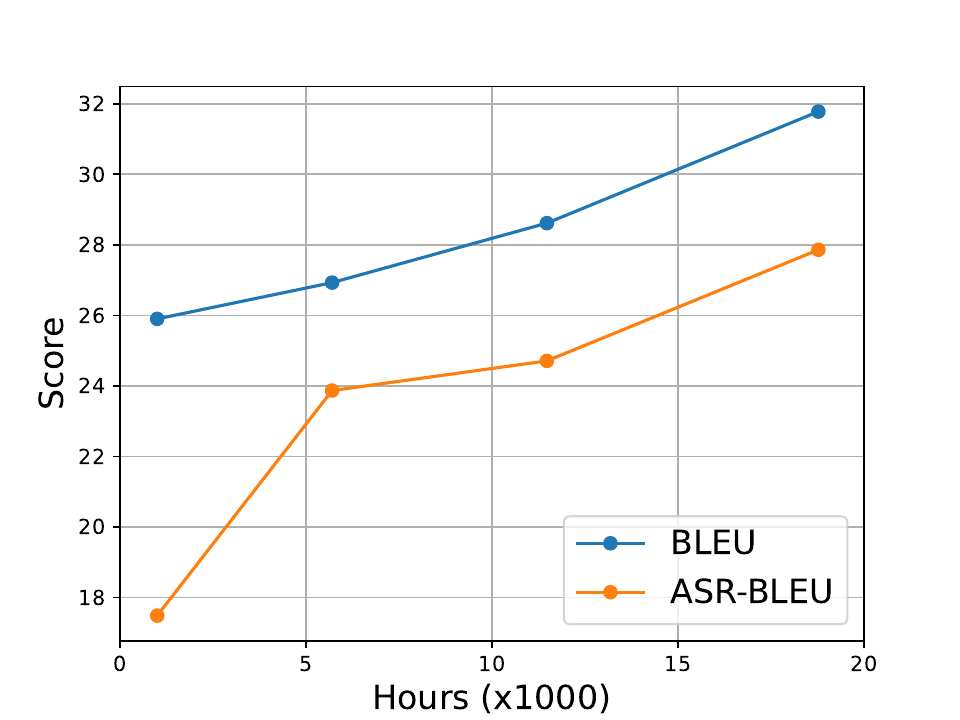}
    \caption{S2TT (BLEU) and S2ST (ASR-BLEU) results on the CVSS-C test set at different training data volume.}
    \label{fig:fr-en-scaling}
    \vspace{-0.2cm}
\end{figure}

Furthermore, we investigated the impact of data scale on final model performance for the FR→EN task. As shown in Figure~\ref{fig:fr-en-scaling}, there is a strong, positive correlation between the amount of monolingual training data (measured in thousands of hours) and the resulting translation quality. Both the text-based BLEU and speech-based ASR-BLEU scores increase consistently as the data volume grows. Together, these experiments confirm that RosettaSpeech not only trains efficiently but also scales effectively with data. While limitations in computational resources and data availability for the source languages prevented a more exhaustive exploration of this scaling potential, our results demonstrate that the model already achieves a powerful level of performance even under the current constraints.



\section{Conclusion}
RosettaSpeech is a novel framework that achieves state-of-the-art, zero-shot speech-to-speech translation using only monolingual data. By leveraging text as an intermediate training bridge, our method bypasses the need for parallel speech corpora and demonstrates robust performance on standard benchmarks for French, Spanish, and German-to-English translation. Ultimately, by removing this critical data dependency, RosettaSpeech provides a scalable and practical blueprint for extending high-fidelity speech translation to the vast number of languages currently underserved by technology.


\section*{Limitations}
Although RosettaSpeech demonstrates promising results, we acknowledge several limitations that present avenues for future research. Our current experiments are confined to translation from a few high-resource European languages into English, and extending the framework to a more diverse set of languages, especially low-resource ones, is a critical next step to assess its broader applicability. Furthermore, the models are currently designed for one-directional, many-to-one translation (into English); a more advanced implementation would support bidirectional or any-to-any translation for greater real-world utility. Our scaling analysis focused exclusively on the size of the training data. A valuable future direction, should sufficient computational resources become available, would be to explore the impact of scaling the model size itself. Finally, for true low-resource languages that lack even basic text translation systems (or where NMT quality is poor), the generated pseudo-labels may suffer from hallucinations or semantic errors, degrading the final S2ST performance. Therefore, our method is best suited for languages that are under-served in the speech domain but adequately supported in the text domain. Future work will investigate the minimum NMT performance threshold required for effective S2ST distillation and explore techniques to mitigate noise in NMT-generated targets.


\newpage
\bibliography{custom}
\newpage
\appendix

\section{Evaluation metrics}\label{app:metric}
\begin{itemize}
    \item \textbf{BLEU}: SacreBLEU~\cite{post2018call} between the translated text and the reference.
    \item \textbf{ASR-BLEU}: Transcribe the generated speech with a pre-trained ASR model~\cite{baevski2020wav2vec}, then compute SacreBLEU~\cite{post2018call} against the reference text.
    \item \textbf{COMET}: Compute COMET~\cite{rei2022comet} with a reference-based model given (source text, hypothesis text, reference text). The score is in \([0,1]\), where higher is better.
    \item \textbf{BLASER 2}: Use the reference-free BLASER-2.0 QE model~\cite{barrault2023seamlessm4t} to score translation quality given (source speech, hypothesis speech) by comparing their SONAR embeddings. Scores are in \([1,5]\) and higher is better.
    \item \textbf{Speaker Similarity (SIM)}: Measured by computing the cosine similarity of speaker embeddings, which are extracted from both the generated and original target speech using a WavLM-based speaker verification model~\cite{chen2022wavlm}. Scores are in \([0, 1]\) and higher is better.
    \item \textbf{NMOS}: Assess the naturalness and intelligibility of the synthesized speech. Scores are in \([1, 5]\) and higher is better.
\end{itemize}



\section{Subjective Evaluation}

To compute our subjective evaluation metrics NMOS, we recruited Amazon Mechanical Turk workers with a minimum approval rate of 99\% and at least 1000 successful HITs. We restricted the worker pool to individuals located in Australia, the United Kingdom, and the United States.

\begin{figure*}[t]
    \centering
    \includegraphics[width=1.0\linewidth]{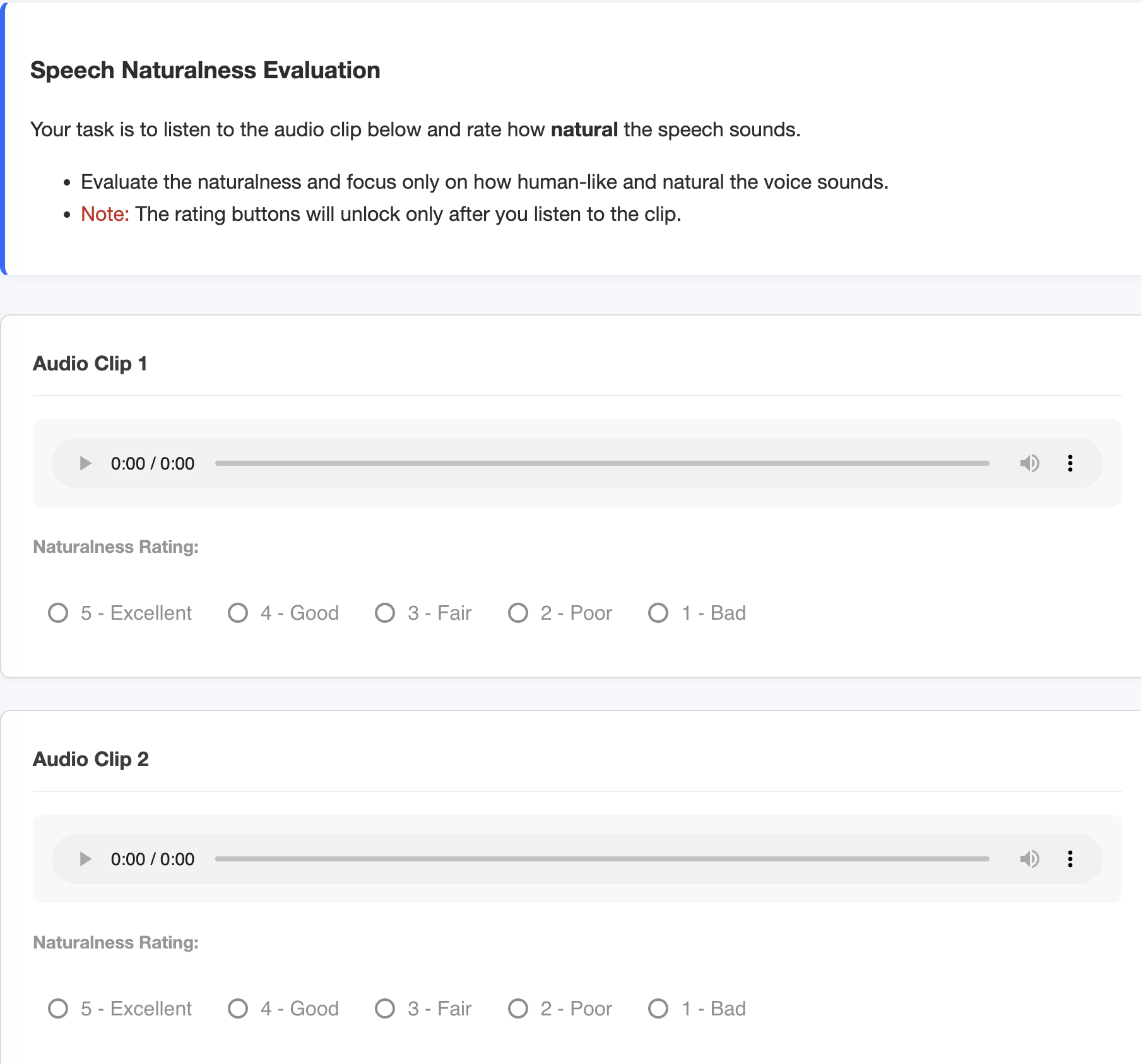}
    \caption{Naturalness Mean Opinion Score (NMOS) Annotation UI}
    \label{fig:nmos}
\end{figure*}

Each sample was annotated by 3 different annotators. We display annotation UIs for our metrics in Figure~\ref{fig:nmos}.

\end{document}